\documentstyle[aps,preprint]{revtex}
\begin{document}
\title{Time-convolutionless reduced-density-operator theory of a 
noisy quantum channel: a two-bit
quantum gate for quantum information processing}
\author{D. Ahn\thanks{Also with Department of Electrical Engineering,
University of Seoul, Seoul 130-743, Korea}\thanks{Electronic address;
dahn@uoscc.uos.ac.kr}, 
J. H. Oh,
K. Kimm, and 
S. W. Hwang\thanks{Permanent address: Department of Electronics Engineering,
 Korea University, 5-1 Anam, Sungbook-ku, Seoul 136-701, Korea}}
\address{Institute of Quantum Information Processing and Systems\\
University of Seoul 90, Jeonnong, Tongdaemoon-ku\\
Seoul 130-743, Republic of Korea}
\maketitle
\begin{abstract}
\baselineskip 16pt
An exact reduced-density-operator for the output quantum states
in time-convolutionless form was derived by solving the quantum Liouville
equation which governs the dynamics of a noisy quantum channel by using a
projection operator method and both advanced and retarded propagators in time.
The formalism developed in this work is general enough to model a noisy
quantum channel provided specific forms of the Hamiltonians for the system,
reservoir, and the mutual interaction between the system and the reservoir are
given.
Then, we apply the formulation to model a two-bit quantum gate composed of
coupled spin systems in which the Heisenberg coupling is controlled by the
tunneling barrier between neighboring quantum dots. Gate Characteristics
including the entropy, fidelity, and purity are calculated numerically for
both mixed and entangled initial states. 
\end{abstract}
\section{Introduction}
There has been a considerable interest in the quantum theory of information and
computation for the past several
years~\cite{Lloyd,Bennett,Jozsa94,Schumacher95,Shor95,Steane,Laflamme,Nielsen,Jozsa95,Wiesner,Brassard,Shor94,DiVincenzo,Barenco,Loss,Schumacher96}. 
Especially, quantum-mechanical
properties of coding~\cite{Jozsa94,Schumacher95}, 
noisy-channels including error-correcting
codes~\cite{Shor95,Steane,Laflamme,Nielsen} and
channel fidelity~\cite{Jozsa95}, and 
computation~\cite{Wiesner,Brassard,Shor94,DiVincenzo,Barenco,Loss} 
have been studied in detail.
It was shown~\cite{DiVincenzo,Barenco} that any quantum 
computation procedure can be decomposed into operations on single-bit
gates and a two-bit gate which involves an entanglement operation on
two quantum bits or qubits. Presence of decoherence and imperfections 
cause the operations of 
these quantum gates away from the ideal ones and as a result one can regard
these gates  as a part of noisy quantum channels. 
Detailed analysis of these channels are necessary for the complete
understanding of general quantum information process.
Mathematically, the dynamics of quantum channels or generalized quantum gates
involves the transformation of input quantum 
states represented by a density
operator $\rho$ into an output states $\rho'$~\cite{Schumacher96}, i.e.,
\begin{eqnarray}
\label{rhodyn}
\rho \buildrel {\cal E}\over\longrightarrow \rho'={\cal E}[\rho],
\end{eqnarray}
where we assume ${\cal E}$ is a linear mapping but is not necessarily a unitary
transformation if one considers an open system interacting with the reservoir
such as noisy quantum channels.
A model of a noisy quantum channel would involve several Hamiltonians for
the system representing qubits, reservoir and the mutual interaction between
the system and the reservoir that causes the decoherence or noise. 
The density operator is then governed by the quantum 
Liouville equation~\cite{Reichl} which
is an integro-differential equations and in general, it is nontrivial to obtain
the solution of the form given by Eq.~(\ref{rhodyn}).
Rather, one is expected to get the solution for
the density operator for the output states in
Volterra type integral equation:
\begin{eqnarray}
\label{volterra}
\rho(t)=A(t,0)\rho(0)+\int d\tau B(t,\tau)\rho(\tau)
\end{eqnarray}
where $A$ is a propagator and $B$ is a memory kernel. In general,
it is very difficult to solve for the memory kernels of the time-convolution
form equation (\ref{volterra}) self-consistently 
and almost always, one must be content
with the narrowing limit or the fast modulation limit~\cite{Saeki82}.

Some time ago, the time-convolutionless equations of motion in the Heisenberg
picture was suggested by Tokuyama and Mori~\cite{Tokuyama} 
to overcome above mentioned
difficulties for problems in nonequilibrium statistical mechanics. 
These formulations were then developed in the Schr{\"o}dinger picture by using
the projection operator 
technique~\cite{Hashitsume,Saeki82,Saeki86}. 
One of the authors applied the
time-convolutionless formulation to the model of quantum devices for detailed
numerical 
study~\cite{Ahn94,Ahn95,Ahn97,Ahn98}. 
It was shown that the time-convolutionless formulation can
also incorporate both non-Markovian relaxation and renormalization of the
memory effects.

Recently, Loss and DiVincenzo~\cite{Loss} 
has made a comprehensive study of the two-bit
quantum gate taking into account the effect of decoherence on the gate
operation using the reduced density operator in the
time-convolution formulation.
Their results indicate  that the detailed analysis of 
the decoherence process is
important for the reliable operation of quantum gates utilizing
controlled, nonequilibrium time evolution of solid-state spin systems.

In order to make the reduce-density operator for the output quantum states
of the form given by  the equation (\ref{rhodyn}),
several approximations including the Born approximation were made in their
theory. In our opinion, it would be more convenient if 
there is a way to get exact
solution for the output density-operator in time-convolutionless form 
given by (\ref{rhodyn}).

In this paper, we first derive the exact solution for the
reduced-density-operator of the output quantum states in time-convolutionless
form by solving the quantum Liouville equation for a quantum channel using the
projection operator method. The formalism we develop would be general enough
to model a realistic quantum channel or a quantum gate. Secondly, we apply the
theory to model a two-bit quantum gate composed of coupled spin systems in
which the Heisenberg coupling is controlled by the tunneling barrier between
neighboring single electron quantum dots.
%%%%%%%%%%%%%%%%%%%%%%%%%%%%%%%%%%%%%%%%%%%%%%%%%%%%%%%%%%%%%%%%%%%%%%%%
\section{Time-convolutionless reduced-density-operator theory of a quantum
system interacting with a reservoir}

In this section, we study the quantum Liouville equation for a quantum system
which corresponds to a quantum channel or a generalized quantum gate to derive
an equation and to solve for a reduced-density-operator of a system coupled to
a reservoir. An interaction between the system and the reservoir leads to
decoherence. The Hamiltonian of the total system is assumed to be
\begin{eqnarray}
H_T(t)=H_S(t)+H_B + H_{\rm int},
\end{eqnarray}
where $H_S(t)$ is the Hamiltonian of the system representing a quantum gate
(or channel), $H_B$ the reservoir and
$H_{\rm int}$ the Hamiltonian for the interaction of the system with its
reservoir. 
The evolution of the system might include a coding, transmission
and decoding process. 
The equation of motion for the density operator $\rho_T(t)$ of the
total system is given by a quantum Liouville equation
\begin{eqnarray}
\frac{d}{dt}\rho_T(t)&=&-i[H_T,\rho_T] \nonumber \\
&=&-i L_T \rho_T,
\label{lioueq}
\end{eqnarray}
where
\begin{eqnarray*}
L_T(t)=L_S(t) +L_B +L_{\rm int}
\end{eqnarray*}
is the Liouville superoperator in one-to-one correspondence with the
Hamiltonian. In  this work, we use a unit where $\hbar=1$. In order to derive
an equation and to solve for a system alone, it is convenient to use the
projection operators~\cite{Nakajiama,Zwanzig} 
which decompose the total system by eliminating the
degrees of freedom for the reservoir. We define time-independent projection
operators $\underline{P}$ and $\underline{Q}$ as~\cite{Hashitsume} 
\begin{eqnarray}
\underline{P} X= \rho_B {\rm tr}_B(X),~~
\underline{Q}=1-\underline{P},
\end{eqnarray}
for any dynamical variable $X$.
Here ${\rm tr}_B$ indicates a partial trace over the quantum reservoir.
Projection operators satisfy the operator identity
$ \underline{P}^2=\underline{P},\underline{Q}^2=\underline{Q}$ and 
$\underline{P}\underline{Q}=\underline{Q}\underline{P}=0$.
The information of the system is then contained in the reduced density
operator $\rho(t)$ which is defined by
\begin{eqnarray}
\rho(t)&=& {\rm tr}_B \rho_T(t)\nonumber \\
&=& {\rm tr}_B \underline{P} \rho_T(t).
\end{eqnarray}
In order to derive a time-convolutionless equation, we first multiply 
Eq.~(\ref{lioueq}) by $\underline{P}$ and $\underline{Q}$ to obtain coupled
equation for $\underline{P}\rho_T(t)$ and $\underline{Q}\rho_T(t)$:
\begin{eqnarray}
\frac{d}{dt}\underline{P}\rho_T(t)
= -i \underline{P}\rho_T\underline{P} \rho_T(t)
+i \underline{P} L_T(t) \underline{Q} \rho_T(t),
\label{peq1}
\end{eqnarray}
\begin{eqnarray}
\frac{d}{dt}\underline{Q}\rho_T(t)
= -i \underline{Q}\rho_T\underline{Q} \rho_T(t)
+i \underline{Q} L_T(t) \underline{P} \rho_T(t).
\label{qeq1}
\end{eqnarray}
We assume that the channel was turned on at $t=0$ and the input state prepared
at $t=0$, $\rho(t=0)$ was isolated with the reservoir at $t=0$, i.e., 
$\underline{Q}\rho_T(0)=0$~\cite{Hashitsume}.

The formal solution of (\ref{qeq1}) is given by~\cite{Ahn94}
\begin{eqnarray}
\label{qsol}
\underline{Q}\rho_T(t)=-i \int_0^t d\tau
\underline{H}(t,\tau)\underline{Q} L_T(\tau) \underline{P}\rho_T(\tau),
\end{eqnarray}
where the projected propagator $\underline{H}(t,\tau)$ of the total system is
given by
\begin{eqnarray}
\underline{H}(t,\tau)=\underline{T}\exp\left\{
-i\int_\tau^t ds \underline{Q}L_T(s)\underline{Q}\right\}.
\end{eqnarray}
Here $\underline{T}$ denotes the time-ordering operator.
Because Eq.~(\ref{qsol}) is in time-convolution form, we transform the memory
kernel in (\ref{qsol}) into time-convolutionless form~\cite{Ahn94} 
by substituting the formal solution of (\ref{lioueq})
\begin{eqnarray}
\label{tsol}
\rho_T(\tau)=\underline{G}(t,\tau)\rho_T(t)
\end{eqnarray}
into Eq.~({\ref{qsol}). The anti-time evolution operator 
$\underline{G}(t,\tau)$ of the total system is defined by
\begin{eqnarray*}
\underline{G}(t,\tau)=\underline{T}^c
\exp\left\{ i \int_\tau^t ds L_T(s)\right\},
\end{eqnarray*}
where $\underline{T}^c$ is the anti-time-ordering operator.
From Eq.~(\ref{qsol}) and (\ref{tsol}), we obtain
\begin{eqnarray}
\label{qtsol}
\underline{Q}\rho_T(t)=\{\theta(t)-1\}\underline{P}\rho_T(t)
\end{eqnarray}
where
\begin{eqnarray}
\theta^{-1}(t)&=&g(t)\nonumber \\
&=&1+i\int_0^t d\tau \underline{H}(t,\tau)\underline{Q}
L_T(\tau)\underline{P}~\underline{G}(t\tau)
\end{eqnarray}
By substituting Eq.~({\ref{qtsol}) into (\ref{peq1}), we obtain the
time-convolutionless equation of motion for $\underline{P}\rho_T(t)$ as
\begin{eqnarray}
\label{peq2}
\frac{d}{dt}\underline{P}\rho_T(t)
=-i\underline{P}L_T(t)\underline{P}\rho_T(t)
-i\underline{P}L_T(t)\{\theta(t)-1\}\underline{P}\rho_T(t)
\end{eqnarray}
It can be shown that the formal solution of (\ref{peq2}) is given by
\begin{eqnarray}
\label{psol}
\underline{P}\rho_T
=\underline{U}(t,0)\underline{P}\rho_T(0)
-i\int_0^t ds \underline{U}(t,s)\underline{P}L_T(s)
\{\theta(s)-1\}\underline{P}\rho_T(s),
\end{eqnarray}
where the projected propagator $\underline{U}(t,\tau)$ of the system is
defined by
\begin{eqnarray}
\underline{U}(t,\tau)
=\underline{T}\exp\left\{
-i\int_0^t ds \underline{P} L_T(s)\underline{P}\right\}.
\end{eqnarray}
To transform Eq.~(\ref{psol}) into time-convolutionless form once again, 
we substitute
\begin{eqnarray}
\rho_T(s)=\underline{G}(t,s)\rho_T(t)
\end{eqnarray}
into (\ref{psol}) to obtain:
\begin{eqnarray}
\underline{P}\rho_T(t)&=&
\underline{U}(t,0)\underline{P}\rho_T(0)
-i\int_0^t ds\underline{U}(t,s)\underline{P}L_T(s)
\{\theta(s)-1\} \underline{P}~\underline{G}(t,s)\rho_T(t)\nonumber\\
&=&\underline{U}(t,0)\underline{P}\rho_T(0)
-i\int_0^t ds\underline{U}(t,s)\underline{P}L_T(s)
\{\theta(s)-1\} \underline{P}~\underline{G}(t,s)\underline{P}\rho_T(t)
\nonumber \\
&&-i\int_0^t ds\underline{U}(t,s)\underline{P}L_T(s)
\{\theta(s)-1\} \underline{P}~\underline{G}(t,s)\underline{Q}\rho_T(t)
\nonumber \\
&=&\underline{U}(t,0)\underline{P}\rho_T(0)
-i\int_0^t ds\underline{U}(t,s)\underline{P}L_T(s)
\{\theta(s)-1\} \underline{P}~\underline{G}(t,s)
\theta(t)\underline{P}\rho_T(t).
\label{psol2}
\end{eqnarray}
By the way,
\begin{eqnarray}
\underline{P}\rho_T(t)&=&
\rho_B{\rm tr}_B \left(\rho_T(t)\right) \nonumber \\
&=&\rho_B\rho(t),
\end{eqnarray}
and
\begin{eqnarray}
\underline{P}L_T(t)\underline{P}&=&
\underline{P}(L_S(t)+L_B+L_{\rm int})\underline{P} \nonumber \\
&=& \underline{P}L_S(t)\underline{P} \nonumber \\
&=&L_S(t)\underline{P}.
\end{eqnarray}
Then
\begin{eqnarray}
\underline{U}(t,0)\underline{P}\rho_T(0)
&=&\underline{T}\exp\left\{
-i \int^t_0 ds \underline{P} L_T(s)\underline{P}\right\}
\underline{P}\rho_T(0) \nonumber \\
&=&\underline{T}\exp\left\{
-i \int^t_0 ds L_S(s)\underline{P}\right\} 
\underline{P}\rho_T(0) \nonumber \\
&=&\underline{U}_S(t,0)\underline{P}\rho_T(0) \nonumber \\
&=&\underline{U}_S(t,0)\rho_B \rho(t).
\label{rel1}
\end{eqnarray}
Here $\underline{U}_S(t,0)$ denotes the propagator of the system.
Likewise,
\begin{eqnarray}
&&\underline{U}(t,s)\underline{P}L_T(s)
\{\theta(s)-1\} \underline{P}~\underline{G}(t,s)
\theta(t)\underline{P}\rho_T(t) \nonumber \\
&=&\underline{U}_S(t,s)\rho_B{\rm tr}_B\Big[
L_T(s)\{\theta(s)-1\}\rho_B{\rm tr}_B\{ \underline{G}(t,s)\theta(t)
\rho_B\}\Big]\rho(t) \nonumber \\
&=&\underline{U}_S(t,s)\rho_B{\rm tr}_B\Big[
L_T(s)\{\theta(s)-1\}\rho_B\Big]
{\rm tr}_B\Big[\underline{G}(t,s)\theta(t)\rho_B\Big]\rho(t).
\label{rel2}
\end{eqnarray}
Substituting (\ref{rel1}) and (\ref{rel2}) into (\ref{psol2}), we obtain
\begin{eqnarray}
\rho(t)&=&\underline{U}_S(t,0)\rho(0)\nonumber \\
&&-i\int_0^t ds \underline{U}_S(t,s) {\rm tr}_B\Big[
L_T(s)\{\theta(s)-1\}\rho_B\Big]
{\rm tr}_B\Big[\underline{G}(t,s)\theta(t)\rho_B\Big]
\rho(t),
\end{eqnarray}
or 
\begin{eqnarray}
\rho(t)&=&{\cal E}(t)\rho(0)\nonumber \\
&=&\underline{W}^{-1}(t)\underline{U}_S(t,0)\rho(0),
\label{rho1}
\end{eqnarray}
with
\begin{eqnarray}
\underline{W}(t)&=&
1+i\int_0^t ds\underline{U}_S(t,s) {\rm tr}_B\Big[
L_T(s)\{\theta(s)-1\}\rho_B\Big]
{\rm tr}_B\Big[\underline{G}(t,s)\theta(t)\rho_B\Big] \nonumber \\
&=&1+i\int_0^t ds\underline{U}_S(t,s) {\rm tr}_B\Big[
L_{\rm int} \Sigma(s)\{1-\Sigma(s)\}^{-1}\rho_B\Big] \nonumber \\
&&\hspace{5em}\times{\rm tr}_B\Big[
\underline{U}_0(s)\underline{R}(t,s)\underline{U}^{-1}_0(t)
\{1-\Sigma(t)\}^{-1}\rho_B\Big].
\label{weq}
\end{eqnarray}
Here, we define
\begin{eqnarray}
\Sigma(t)=1-\theta^{-1}(t),
\end{eqnarray}
\begin{eqnarray}
\underline{U}_0(t)=e^{-it L_B}\underline{U}_S(t),
\end{eqnarray}
and
\begin{eqnarray}
\underline{R}(t,\tau)=
\underline{T}^c \exp\left\{
i\int^t_\tau ds \underline{U}_0^{-1}(s)L_{\rm int} \underline{U}_0\right\},
\end{eqnarray}
where $\underline{U}_0(t)$ is the evolution operator of the system with the
reservoir and $\underline{R}(t,\tau)$ is 
the evolution operator~\cite{Saeki86} of the total
system in the interacting picture. In (\ref{weq}), we use the identities
$\underline{P}L_T(s)\underline{Q}=\underline{P}L_{\rm int}\underline{Q}$
and $\underline{H}(t,\tau)\underline{Q}=\underline{Q}~\underline{H}(t,\tau)$.

Detailed expression for $\Sigma(t)$ becomes
\begin{eqnarray}
\Sigma(t)&=&1-\theta^{-1}(t) \nonumber \\
&=& -i \int^t_0 d\tau \underline{H}(t,\tau)\underline{Q}
L_T(\tau)\underline{P} ~\underline{G}(t,\tau) \nonumber \\
&=&-i \int^t_0 d\tau \underline{H}(t,\tau)\underline{Q}
L_{\rm int}(\tau)\underline{P} ~\underline{G}(t,\tau) \nonumber \\
&=&-i \int^t_0 d\tau
\underline{U}_0(t)\underline{S}(t,\tau)\underline{U}_0^{-1}
\underline{Q}L_{\rm int}\underline{P}~\underline{U}_0(\tau)
\underline{R}(t,\tau)\underline{U}_0^{-1}(t),
\end{eqnarray}
with
\begin{eqnarray}
\underline{S}(t,\tau)=\underline{T}\exp\left\{
-i \int^t_\tau ds ~\underline{Q}\underline{U}_0^{-1}(s)
L_{\rm int} \underline{U}_0(s) \underline{Q}\right\},
\end{eqnarray}
where $\underline{S}(t,\tau)$ is the 
projected propagator~\cite{Saeki86} of the total system
in the interaction picture. It is now obvious from (\ref{rho1}) and
(\ref{weq}), the exact solution $\rho(t)$ for the output quantum state is 
in time-convolutionless form
given by Eq.~(\ref{rhodyn}) which is employed in the description of
quantum information processing and computation~\cite{Schumacher96}. 

We now consider the case when the system is interacting weakly with the
reservoir and expand (\ref{weq}) up to the second order in powers of the
interaction Hamiltonian $H_{\rm int}$. The renormalization of the unperturbed
energy of the system and the first order of the interaction $H_{\rm int}$ 
gives~\cite{Hashitsume,Saeki82,Saeki86}
\begin{eqnarray}
\underline{P}L_{\rm int}\underline{P}=0.
\end{eqnarray}
Then in the lowest order Born approximation which is valid up to the order
$(H_{\rm int})^2$, we obtain
\begin{eqnarray}
\underline{W}^{(2)}(t)&=& 
1+i \int^t_0 ds \underline{U}_S(t,s){\rm tr}_B\Big[
L_{\rm int}\Sigma^{(1)}(s)\rho_B\Big]{\rm tr}_B\Big[
\underline{U}_0(s) \underline{U}_0^{-1}(t)\rho_B\Big]\nonumber \\
&=& 1+i \int^t_0 ds \underline{U}_S(t,s){\rm tr}_B\Big[
L_{\rm int}\Sigma^{(1)}(s)\rho_B\Big]
\underline{U}_S^{-1}(t,s),
\label{w2}
\end{eqnarray}
or
\begin{eqnarray}
\Big[\underline{W}^{(2)}(t)\Big]^{-1}=
 1-i \int^t_0 ds \underline{U}_S(t,s){\rm tr}_B\Big[
L_{\rm int}\Sigma^{(1)}(s)\rho_B\Big]
\underline{U}_S^{-1}(t,s),
\end{eqnarray}
and 
\begin{eqnarray}
\label{Mequation}
\underline{{\cal E}}^{(2)}=\left[
1-i\int^t_0 ds \underline{U}_S(t,s){\rm tr}_B\Big[
L_{\rm int}\Sigma^{(1)}(s)\rho_B\Big]
\underline{U}_S^{-1}(t,s)\right]\underline{U}_S(t,0)
\end{eqnarray}
Here
\begin{eqnarray}
\Sigma^{(1)}(s)&=&-i \int^s_0 d\tau 
\underline{U}_0(s)\underline{U}_0^{-1}(\tau)
\underline{Q}L_{\rm int}\underline{P}~\underline{U}_0(\tau)
\underline{U}_0^{-1}(s) \nonumber \\
&=&-i\int^s_0 d\tau \underline{U}_0(s,\tau)L_{\rm
int}\underline{U}_0^{-1}(s,\tau) .
\label{sigma1}
\end{eqnarray}
The time-convolutionless form of the output reduced-density-operator
\begin{eqnarray}
\rho(t)=\underline{{\cal E}}^{(2)}(t)\rho(0)
\end{eqnarray}
together with (\ref{w2})-(\ref{sigma1}) can be used in any time scale and 
is valid up to the second order in powers in the interaction between the
system and the reservoir.

In the next section, reduced-density-operator for the output quantum state is
used to study the two-bit quantum gate utilizing coupled spin system in
nonequilibrium situation.

%%%%%%%%%%%%%%%%%%%%%%%%%%%%%%%%%%%%%%%%%%%%%%%%%%%%%%%%%%%%%%%%%%%%%%%%
\section{Decoherence of two-bit quantum gate}

We consider a two-bit quantum gate based on nonequilibrium dynamics of 
 the spin of excess electrons in
quantum dots~\cite{Loss}. In this system, the gate 
operation is controlled by
an electrical tunneling between two quantum dots. Projecting out the 
spatial parts
of wavefunctions of electrons, we model the system by 
the Hubbard Hamiltonian~\cite{Hubb};
\begin{eqnarray}
H_S(t) = J(t) \vec{S_1}\cdot\vec{S_2}
\label{hubbard}
\end{eqnarray}
where $J(t)$ is time-dependent Heisenberg coupling which involves
the energy difference between the spin singlet 
and triplet states.
If we turn on $J(t)$ for $\int dt J(t)=J_0 \tau_s = \pi$,
the unitary operator associated with the Hamiltonian (\ref{hubbard})
gives the swap operation up to overall phase difference; 
if $|i,j\rangle$ labels the spin states of two electrons
in the $S_z$ basis with $i,j=\uparrow,\downarrow$, then swap operation
$U_{\rm swap}$ on two registers $|i,j\rangle$ gives 
$U_{\rm swap} |i,j\rangle=|j,i\rangle$.

In reality, quantum-dot system of our interest is not a closed system, so we
have to take into account of  the decoherence effects 
due to the interaction with
the environment which is coupled with the system.
{}For the action of the environment during the gate operation, we use
a Calderia-Leggett-type model~\cite{Loss} 
where a set of harmonic oscillators are coupled
linearly to the system spins by
\begin{eqnarray}
\label{int}
H_{\rm int}=\lambda(\vec{S_1}\cdot\vec{b_1}+\vec{S_2}\cdot\vec{b_2})
\end{eqnarray}
Here, $b^j_i=\sum_{\alpha} g_{\alpha}(a_{\alpha,i}^j
+{a^j_{\alpha,i}}^{\dagger})$ is a
fluctuating quantum field whose unperturbed motion is governed 
by the harmonic-oscillator
Hamiltonian,
\begin{eqnarray}
\label{bath}
H_B(t) = \sum_{\alpha} \omega_\alpha a^{\dagger}_{\alpha} a_{\alpha}
\end{eqnarray}
where $a^{\dagger}_{\alpha}$ $(a_{\alpha})$ are bosonic creation (annihilation)
operator and $\omega_\alpha$ are the corresponding frequencies with spectral
distribution function 
$A(\omega)=\pi\sum_\alpha g_\alpha^2 \delta(\omega-\omega_\alpha)$.

For a coupled spin system, the 
evolution operator ${\cal E}^{(2)}$ given by 
Eq. (\ref{Mequation}) can be written down explicitly
in terms of spin operators.
Substituting (\ref{hubbard})-(\ref{bath}) into definitions 
for $\underline{U}_0$ 
and $L_{\rm int}$, the integrand
of Eq. (\ref{Mequation}) can be written as,
\begin{eqnarray}
\label{trb}
 & & {\rm tr}_B \left[ L_{\rm int}\Sigma^{(1)}(s)\rho_B\right]
\underline{U}_S^{-1}(t,s)
\underline{U}_S(t,0)\rho(0)     \nonumber \\
&=& -i\int_0^s d\tau {\rm tr}_B\left[ L_{\rm int}
\underline{U}_0(s,\tau)L_{\rm int} \underline{U}_0^{-1}(s,\tau) \rho_B\right]
 \underline{U}_S(s,0)\rho(0) \nonumber \\
&=& -i\lambda^2 \sum_{ijkl}\int_0^s d\tau \Big\{
 [ S^j_i,S^l_k(\tau-s) (\underline{U}_S(s,0)\rho(0)) ]
{\rm tr_B}\{b_i^j b_k^l(\tau-s)\rho_B\} \nonumber \\
& &~~~~~+[(\underline{U}_S(s,0)\rho(0)) S^l_k(\tau-s),S^j_i]
{\rm tr_B}\{b_k^l(\tau-s) b_i^j \rho_B\} \Big\} \nonumber \\
&=&-i\sum_{ij}\int_0^sd\tau \Big\{[ S^j_i,S^j_i(\tau-s) 
(\underline{U}_S(s,0)\rho(0)) ]\{\Gamma(\tau-s)-i\Delta(\tau-s)\}
\nonumber \\
& &~~~~~+[(\underline{U}_S(s,0)\rho(0)) S^j_i(\tau-s),S^j_i]
\{\Gamma(\tau-s)+i\Delta(\tau-s)\} \Big\},
\end{eqnarray}
where the trace over the heat bath is done for  
the harmonic oscillator eigenstates,
\begin{eqnarray}
\label{gamma}
{\rm Tr_B}\{b_k^l(t)b_i^j \rho_B\} =
\delta_{ik}\delta_{jl}\frac{1}{\pi}\int_0^\infty A(\omega)
\left\{e^{-i\omega t}+
\frac{2\cos(\omega t)}{e^{\omega/k_BT}-1} \right\} d\omega,
\end{eqnarray}
and we define $\Gamma(t)$ and $\Delta(t)$ as
\begin{eqnarray}
\Gamma(t)+i\Delta(t) = \lambda^2 {\rm Tr_B}\{b_i^j(t)b_i^j \rho_B\}.
\end{eqnarray}
Then, Eq. (\ref{Mequation}) leads to
\begin{eqnarray}
\label{Dequation}
\underline{{\cal E}}^{(2)} &=& \underline{U}_S(t,0) \bigg[
1-\int^t_0 ds \int^s_0 d\tau
\sum_{ij}\Big\{ [S_i^j(s),S^j_i(\tau)\rho(0)]\big\{
\Gamma(\tau-s)-i\Delta(\tau-s)\big\} 
\nonumber \\
&+&   [\rho(0) S_i^j(\tau),S^j_i(s)]\big\{\Gamma(\tau-s)+i\Delta(\tau-s)\big\}
 \Big\} \bigg].
\end{eqnarray}

Now we evaluate the density operator in basis representation;
$\rho(t)=\sum_{\alpha\beta}\rho_{\alpha\beta}(t) e_{\alpha\beta}$,
$e_{\alpha\beta}$ is the basis for the density operators, and in this work
we choose $e_{\alpha\beta}$ as the multiplet states, $i.e.$
$e_{\alpha\beta}=|\alpha\rangle\langle\beta|$
with $\alpha,\beta=1,2,3,4$;
$| 1\rangle=|\uparrow\uparrow\rangle,
 | 2\rangle=(|\uparrow\downarrow\rangle
+|\downarrow\uparrow\rangle)/\sqrt{2},
 | 3\rangle=|\downarrow\downarrow\rangle$, and 
$| 4\rangle=(|\uparrow\downarrow\rangle
-|\downarrow\uparrow\rangle)/\sqrt{2}$.
By defining the inner product like $(e_{\alpha\beta}, e_{\gamma\delta})
={\rm tr}[e_{\alpha\beta}^\dag e_{\gamma\delta}]
=\delta_{\alpha\beta}\delta_{\gamma\delta}$, 
$\rho_{\alpha\beta}(t)$ is obtained as;
\begin{eqnarray}
\label{rhocomp}
\rho_{\alpha\beta}& =& (e_{\alpha\beta},\rho(t)) \nonumber \\
       &=& (e_{\alpha\beta},{\cal E}^{(2)} \rho(0))=
\sum_{\gamma\delta}(e_{\alpha\beta},{\cal E}^{(2)}e_{\gamma\delta})
\rho(0)_{\gamma\delta}
=\sum_{\gamma\delta}{\cal E}^{(2)}_{\alpha\beta|\gamma\delta}
 \rho(0)_{\gamma\delta}
\end{eqnarray}
where $\rho(0)_{\gamma\delta}$ expansion coefficients 
of the initial density operator.
Without the interaction with environment, 
$i.e.$ the absence of the second term in Eq.~(\ref{Dequation}),
${\cal E}^{(2)}_{\alpha\beta|\gamma\delta}$ is reduced to
${U_S(t)}_{\alpha\beta|\gamma\delta}$ and evaluated 
on the multiplet basis as
\begin{eqnarray}
(e_{\alpha\beta},U_S(t)e_{\gamma\delta})
=\delta_{\alpha\beta}\delta_{\gamma\delta}
e^{-i\overline{t}(E_{\alpha}-E_\beta)},
\label{urep}
\end{eqnarray}
where $E_{1,2,3}=J_0/4$ and $E_4=-3J_0/4$ are the triplet 
and singlet energy eigenvalues.
Here, $\overline{t}$ has its value $t$ ($\tau_s$) 
if $t$ is less(larger) than $\tau_s$.
Then, $\underline{U}_S(t)$ becomes the swap operator,
$U_{S}(t)=e^{-i\pi/4}\underline{U}_{\rm swap}$ if $t=\tau_s$.

In order to evaluate ${\cal E}^{(2)}$, 
we first calculate the following matrix elements:
\begin{eqnarray}
\label{S1}
(e_{\alpha\beta},\sum_{ij}[S^j_i(s),S^j_i(\tau)e_{\gamma\delta}]) &=&
\sum_{ij}\{ \langle\alpha| S^j_i(s)S^j_i(\tau)
|\gamma\rangle\langle\delta|\beta\rangle\nonumber \\
&-& \langle\alpha| S^j_i(\tau)|\gamma\rangle\langle\delta| S^j_i(s)
|\beta\rangle \}\nonumber \\
&=&\delta_{\delta\beta}\sum_\kappa
 M_{\alpha\kappa\kappa\gamma}
      e^{i\overline{\tau}\omega_{\kappa\gamma}
+i\overline{s}\omega_{\alpha\kappa}}
-M_{\alpha\gamma\delta\beta}
      e^{i\overline{\tau}\omega_{\alpha\gamma}
+i\overline{s}\omega_{\delta\beta  }}
\end{eqnarray}
and
\begin{eqnarray}
\label{S2}
(e_{\alpha\beta},\sum_{ij}[e_{\gamma\delta}S^j_i(\tau),S^j_i(s)]) &=&
\sum_{ij}\{ \langle\delta| S^j_i(\tau)S^j_i(s)
|\beta\rangle\langle\alpha|\gamma\rangle\nonumber \\
&-& \langle\alpha| S^j_i(s)|\gamma\rangle\langle\delta
| S^j_i(\tau)|\beta\rangle \}\nonumber \\
&=&\delta_{\alpha\gamma}\sum_{\kappa }M_{\delta\kappa \kappa \beta}
      e^{i\overline{\tau}\omega_{\delta\kappa}
+i\overline{s}\omega_{\kappa\beta}}
-M_{\alpha\gamma\delta\beta}
      e^{i\overline{\tau}\omega_{\delta\beta}
+i\overline{s}\omega_{\alpha\gamma}}
\end{eqnarray}
where $M_{\alpha\beta\gamma\delta}=\sum_{ij}\langle\alpha| S^j_i
|\beta\rangle\langle\gamma| S^j_i|\delta\rangle$,
$\omega_{\alpha\beta}=E_\alpha-E_\beta$, and
\begin{eqnarray*}
\overline{\tau}(\overline{s})=
\left\{\begin{array}{ll}
\tau(s) & \mbox{if }  \tau(s) < \tau_s\\
\tau_s  & \mbox{otherwise.} 
\end{array}\right.
\end{eqnarray*}
Then, the matrix element of the evolution operator,
${\cal E}^{(2)}_{\alpha\beta|\gamma\delta}$ is obtained 
by substituting (\ref{urep})-(\ref{S2}) into (\ref{Dequation}),
\begin{eqnarray}
{\cal E}^{(2)}_{\alpha\beta|\gamma\delta}
&=& e^{-i\overline{t}\omega_{\alpha\gamma}}
\Big[     \delta_{\alpha\gamma}\delta_{\beta\delta}
    -\delta_{\beta\delta} \sum_\kappa M_{\alpha\kappa\kappa\gamma}
                              p_{\kappa\kappa|\gamma\alpha}(t)
        -\delta_{\alpha\gamma}\sum_\kappa M_{\delta\kappa\kappa\beta}
                              p^*_{\delta\kappa|\gamma\beta}(t)  \nonumber \\
&&\hspace{4em} +M_{\alpha\gamma\delta\beta}
\{p_{\alpha\beta|\gamma\delta}(t)+p^*_{\beta\alpha|\delta\gamma}(t)\}\Big]
\end{eqnarray}
with the time-dependent term $p_{\alpha\beta|\gamma\delta}(t)$ defined by
\begin{eqnarray}
p_{\alpha\beta|\gamma\delta}(t)
=\int^t_0 ds e^{-i\overline{s}\omega_{\beta\delta}}
               \int^s_0 d\tau e^{i\overline{\tau}\omega_{\alpha\gamma}}
                 \{\Gamma(\tau-s)-i\Delta(\tau-s) \}.
\end{eqnarray}
For numerical calculations, it is more convenient to split 
the time integrals of the matrix
$p_{\alpha\beta |\gamma\delta}(t)$ into three parts;
\begin{eqnarray}
p_{\alpha\beta|\gamma\delta}(t)
&=&\int^{\tau_s}_0 ds e^{-is\omega_{\beta\delta}}
                  \int^{s}_0 d\tau e^{i\tau\omega_{\alpha\gamma}}
           \{\Gamma(\tau-s)-i\Delta(\tau-s) \}
\nonumber \\
            & &+\int_{\tau_s}^t ds e^{-i\tau_s\omega_{\beta\delta}}
             \int^{\tau_s}_0 d\tau e^{i\tau\omega_{\alpha\gamma}}
                               \{\Gamma(\tau-s)-i\Delta(\tau-s) \}
\nonumber \\
              & &+\int_{\tau_s}^t ds e^{-i\tau_s\omega_{\beta\delta}}
                \int^{s}_{\tau_s} d\tau e^{i\tau_s\omega_{\alpha\gamma}}
                               \{\Gamma(\tau-s)-i\Delta(\tau-s) \}
\nonumber \\
      &=&\int^{\tau_s}_0 ds e^{is(\omega_{\delta\beta}+\omega_{\alpha\gamma})}
                            \int^{s}_0 d\tau e^{i\tau\omega_{\gamma\alpha}}
                             \{\Gamma(\tau)+i\Delta(\tau)\}
\nonumber \\
                & & +e^{i\tau_s\omega_{\delta\beta}}
                \int_{\tau_s}^t ds e^{is\omega_{\alpha\gamma}}
                \int^{s}_{s-\tau_s} d\tau e^{i\tau\omega_{\gamma\alpha}}
                               \{\Gamma(\tau)+i\Delta(\tau) \}
\nonumber \\
           & & +e^{i\tau_s(\omega_{\delta\beta}+\omega_{\alpha\gamma})}
                               \int_{\tau_s}^t ds
                               \int^{s-\tau_s}_0 d\tau 
                               \{\Gamma(\tau)+i\Delta(\tau) \}.
\label{peq}
\end{eqnarray}
In order to investigate  the dynamics of the density 
operator in non-equilibrium situation, we calculate
Eqs.~(\ref{rhocomp})-(\ref{peq}) numerically,
assuming an Ohmic damping for spectral distribution 
function $A(\omega)=\eta\omega$ with a cutoff frequency $\omega_c$\cite{Palma}.
%%%%%%%%%%%%%%%%%%%%%%%%%%%%%%%%%%%%%%%%%%%%%%%%%%%%%%%%%%%%%%%%%%%%%%%%
%%%%%%%%%%%%%%%%%%%%%%%%%%%%%%%%%%%%%%%%%%%%%%%%%%%%%%%%%%%%%%%%%%%%%%%%
\section{Numerical Results and Discussions}
%%%%%%%%%%%%%%%%%%%%%%%%%%%%%%%%%%%%%%%%%%%%%%%%%%%%%%%%%%%%%%%%%%%%%%%%
We now study dynamics of the density operator for various initial states.
First, we calculate the evolution of the spin states 
during the swap gate operation and
compare our results with those obtained by Loss and DiVincenzo\cite{Loss}. The
initial spin state is chosen to be the spin-up for the second electron while
the first electron is unpolarized;
$\rho(0)=(|\uparrow\uparrow\rangle\langle\uparrow\uparrow|
+|\downarrow\uparrow\rangle\langle\downarrow\uparrow| )/2$.
In the multiplet basis, the initial state is expanded as;
\begin{eqnarray}
\label{irho}
\rho(0) =  \frac{1}{2}| 1\rangle\langle 1| 
         +\frac{1}{4}| 2\rangle\langle 2|
         -\frac{1}{4}| 2\rangle\langle 4|
         -\frac{1}{4}| 4\rangle\langle 2|
         +\frac{1}{4}| 4\rangle\langle 4| .
\end{eqnarray}

Fig.~1-(a) shows the spin polarization calculated using
parameters $\lambda^2\eta=1.8\times 10^{-5}$, $k_BT=300~{\rm K}$, 
$\omega_c=400~{\rm K}$, and $J_0=1~{\rm K}$(solid lines). For 
the interval, $0\le t \le\tau_s$ the 
spin polarization of the first
electron $s=2\langle S_z^1\rangle=2 {\rm tr}[\rho(t) S_z^1]$ 
changes to  nearly a unity whereas
the spin state of the second electron becomes zero(dashed line), 
demonstrating the feasibility of the swap operation.
However, due to the decoherence, we find  that 
a perfect swap operation cannot be achievable. 
In addition, the perturbing fields 
cause the monotonic decreases of the spin polarization with the elapse time
after completion of swap operation.
This means that spin states are becoming thermalized 
owing to the
interaction with the environment, 
which shows the decoherence of the states. 
The decoherence would be  a fundamental problem 
in making a reliable quantum logic gate,
which puts severe restriction on building the realistic quantum computer. 
However, there are several quantum error-correction techniques 
which can compensate imperfections 
introduced by the decoherence during and after  the gate 
operation~\cite{Shor95,Steane,Laflamme,Nielsen}.
Comparing with the result obtained in the 
previous work~\cite{Loss}(dotted line),
 we find that
both calculations yield similar results for $t>\tau_s$ 
except for the value at $t=\tau_s$.
We think that the discrepancy at $t=\tau_s$ is resulted from 
somewhat simplified evaluation of 
the evolution operator in the reference~\cite{Loss}
when the swap operation occurs.

In Fig. 1-(b) and (c), we plot the gate fidelity ${\cal F}$ and 
gate purity ${\cal P}$  which 
characterize the intrinsic properties of the gate,
and are defined as~\cite{Poyatos};
\begin{eqnarray}
{\cal F}
&=& \overline{ \langle \psi_0| U_S^{\dagger}(\bar{t})\rho(t)
                        | \psi_0\rangle }
  =  \frac{1}{6}+\frac{1}{24}\left[\sum_{\alpha}
     {\cal E}^{(2)}_{\alpha\alpha|\alpha\alpha}
   +\sum_{\alpha,\beta} {\cal E}^{(2)}_{\alpha\beta|\alpha\beta}
          e^{i \bar{t}\omega_{\alpha\beta}}
  \right ], 
\label{fidelity}\\
{\cal P} &=& \overline{{\rm tr}[\rho(t)]^2 }
  =  \frac{1}{24}\sum_{\alpha,\beta,\gamma} 
\left[| {\cal E}^{(2)}_{\alpha\beta|\gamma\gamma}|^2
  +\sum_{\delta}\left({\cal E}^{(2)}_{\alpha\beta|\gamma\gamma}
{\cal E}^{(2)*}_{\alpha\beta|\delta\delta}
  +| {\cal E}^{(2)}_{\alpha\beta|\gamma\delta}|^2 \right)
  \right]
\label{purity}
\end{eqnarray}
where the overbar means an average over all possible 
initial state $|\psi_0\rangle$ and $U_S(\bar t)$ is an ideal gate operation
which was turned on during the time interval, $0\le t\le \tau_s$.
The last equalities in 
({\ref{fidelity}) and (\ref{purity}) were derived under the condition
of both trace and hermiticity of 
${\cal E}^{(2)}$ being preserved within our approximation scheme.
For an  ideal quantum gate, the gate fidelity ${\cal F}$  and
the gate purity ${\cal P}$ must be equal to one during the gate operation
because in that case the evolution operator is unitary. 
Our calculation shows that both ${\cal F}$ and ${\cal P}$ 
are found to decrease  almostly
linearly as time elapses, which indicates clearly the presence of 
decoherence effect. As the case of the spin polarization,
the decreasing rates for ${\cal F}$ and ${\cal P}$ are close to 
those obtained in Ref.~\cite{Loss}, 
however its value at $t=\tau_s$ are different 
and our results show more severe
decoherence of the spin state for the same parameters.

Another interesting property of the two-bit gates is 
the von Neumann entropy $\Lambda$ 
defined as $\Lambda=-{\rm tr}[\rho(t)\log_2\rho(t)]$ of a quantum state. 
In Fig.~1-(d), the calculated von Neumann entropy of the spin system
is plotted.
For the initial density operator of Eq. (\ref{irho}), 
its entropy is $\Lambda=1~({\rm bit})$ because
the eigenvalues of $\rho(0)$ are $\{0,0,1/2,1/2\}$. 
As time goes on, the entropy becomes
larger because the thermalization makes the system 
reside equally in all states.
Eventually, the entropy will reach to the maximum value of 
$\Lambda=2~({\rm bits})$ where all four states are equally probable.

To examine the effect of the perturbing field on an entangled state, 
we now consider
a different initial density operator. We assume that the system is in a 
pure spin singlet at $t=0$;
$|\psi_0\rangle
=(|\uparrow\downarrow\rangle-|\downarrow\uparrow\rangle)/\sqrt{2}$
and its density operator
is $\rho(0) = |\psi_0\rangle\langle \psi_0|$. 
In Fig.~2-(a), we plot the diagonal 
components of the density operators in the multiplet basis
as a function of time. $\rho_{44}$ (solid line) 
loses its coherence
linearly to time while other components 
$\rho_{\alpha\alpha}$ grows  as time
elapses. This behavior gives rise to an increasing 
value of the entropy as shown in Fig.~2-(d).
For the pure initial state one can calculate 
the fidelity of the gate without too much difficulty.
We compare fidelity ($\langle\psi_0 |U^{\dagger}_S(\bar{t})
\rho(t)|\psi_0\rangle$) of a given entangled pure state (dotted line) with 
the gate fidelity (solid line) in Fig.~2-(c) and in addition the 
purity (${\rm tr}\rho(t)^2$)  of a
given initial entangled state (dotted line) with gate purity (solid line) in 
Fig.~2-(d). 
In both quantities, there are a slight difference between the cases. 
This implies that although the gate fidelity ${\cal F}$ and gate purity 
${\cal P}$ define the global  characteristics of gate, fidelity and purity
of the gate for a specific input state depends on input itself.
 
Now, we discuss the strength of the decoherence 
which depends on $\Gamma(t)$ and $\Delta(t)$
of Eq. (\ref{Dequation});
\begin{eqnarray}
\Gamma(t)+i\Delta(t)
=\frac{\lambda^2\eta}{\pi}\int_0^{\omega_c} \omega\cos{\omega t}
\coth\left[\frac{\omega}{2k_BT}\right] d\omega
-i\frac{\lambda^2\eta}{\pi}\int_0^{\omega_c} \omega\sin{\omega t} d\omega.
\end{eqnarray}
For a sufficiently high temperature 
$k_BT\gg\omega_c/2$, $\Gamma(t)$ and $\Delta(t)$
are further simplified to 
\begin{eqnarray}
\Gamma(t)+i\Delta(t)=\frac{2\Gamma_0}{\pi\tau_s}\frac{\sin{\omega_c t}}{t}
-i\frac{\Delta_0}{\tau_s}\left[ \frac{\sin{\omega_c t}}{\omega_c t^2}-
\frac{\cos{\omega_c t}}{t} \right]
\end{eqnarray}
with $\Gamma_0=\lambda^2\eta k_BT \tau_s$ 
and $\Delta_0=\lambda^2\eta\omega_c\tau_s/\pi$.
Since a typical value of $\tau_s$ is  $25{\rm psec}$ for $J_0=1K$ and,
thus $\omega_c\tau_s\gg1$, $\Gamma(t)$ and $\Delta(t)$ 
are rapidly oscillating functions.
This implies that the dominant contribution to the decoherence
can be written as 
$\Gamma(t)+i\Delta(t)=2\Gamma_0\delta(t)/\tau_s$ 
in the limit of $\omega_c\tau_s\gg1$.
In this approximation, we find that $p_{\alpha\beta|\gamma\delta}(t)$ 
of Eq. (\ref{peq})
is proportional to $\Gamma_0 t$. This behavior is 
attributed to a linear dependence of
various quantities($s,{\cal F},{\cal P}$) on time.
In addition, we expect that the degradation of the spin 
polarization is also proportional
to $\Gamma_0t$. For this, we examine the evolution of the spin polarization
of the first electron for the initial density operator 
of Eq. (\ref{irho}) for various values
of the coupling constant, $\lambda^2\eta$, and plot results in Fig. 3-(a).
As $\lambda^2\eta$ increases, we find that
more strong decoherence occurs in spin states and its dependence is linear
on $\lambda^2\eta$ as shown in Fig. 3-(b). 
This linear dependence also appears
in the fidelity and purity.

In summary, we first derive an exact reduced-density-operator for the output
quantum states in time-convolutionless form by solving the quantum Liouville
equation for a noisy quantum channel. The formalism developed in this paper
would be general enough to model a noisy quantum channel if various
Hamiltonians for a channel dynamics, environment and an interaction are given.
Secondly, we calculated various characteristics 
including the fidelity, purity, and the change of entropy of
a two-bit quantum gate which is based on the spin exchange
interaction between two quantum dots.  
Our calculation shows it is really important to control the decoherence in
the quantum gate to protect quantum information against corruption.
The decoherence in the quantum logic gate which is extremely
sensitive to it may be a major obstacle to building 
the realistic quantum computer, however, 
it it is expected  that 
as long as the error rate is below some threshold value, a quantum
computer which can give arbitrary accurate answer can be built 
with a reasonable model of decoherence.
In this respect, it will be interesting to investigate the 
implementation of  quantum error correction
technique for this model.
Another interesting study on the present model is to find 
an operator sum representation for the evolution 
operator ${\cal E}$:
\begin{eqnarray}
{\cal E}[\rho]=\sum_{\mu} A_\mu \rho A_\mu^\dagger,
\end{eqnarray}
where $A_\mu$ is an operator acting on the system alone. 
With the operator sum representation, we can calculate various information
theoretical quantities such as the coherent information, entropy exchange, and the
channel capacity~\cite{Schumacher96}. 
We would like to leave this subject for future work.
\acknowledgments{
We thank to  Dr. Ki Jeong Kong and Dr. Jinsoo Kim for valuable discussions.
This work was supported by the Korean Ministry of Science and Technology
through the Creative Research Initiatives Program
under Contract No. 98-CR-01-01-A-08.}

%%%%%%%%%%%%%%%%%%%%%%%%%%%%%%%%%%%%%%%%%%%%%%%%%%%%%%%%%%%%%%%%%%%%%%%%%%%%

\newpage
\begin{center}
{\large Figure Captions}
\end{center}
{\bf Fig}. 1. The calculated spin polarization($s$), fidelity(${\cal F}$),
purity(${\cal P}$), and entropy($\Lambda$) are plotted 
as a function of time(solid lines),
and compared with those obtained in the Ref.~\cite{Loss} (dotted lines).
We assume that the first electron is un-polarized on the initial state
with the second polarized upward. For $0\leq t\leq\tau_s$, 
the swap operation is made
by turning on $J(t)$ and, then, $J(t)=0$.

\hspace{0.3cm}
\newline
\hspace{0.3cm}
{\bf Fig}. 2. For the initial density operator $\rho_{44}$,
we show diagonal components of the density operator as time elapses in (a). 
$\rho_{44}$ decreases monotonically(solid line) whereas others of diagonal 
components become larger(dotted lines).
In (b) and (c), the fidelity and purity are shown 
concerning with(solid line) and
without(dotted line) an average over initial states. 
The evolution of the entropy as plotted
in (d) starts from zero because the initial state is pure. 

\hspace{0.3cm}
\newline
\hspace{0.3cm}
{\bf Fig}. 3. For various values of the coupling constants,
$\lambda^2\eta=0.5\times 10^{-5}$(dotted),
$1.8\times 10^{-5}$(solid), and $3.0\times 10^{-5}$(dashed),
 we show the evolution of the spin polarization
of the first electron in (a) for the initial density of 
Eq. (\ref{irho}). In (b), the degradation
of the spin polarization$(s)$, fidelity(${\cal F}$), 
purity(${\cal P}$), and entropy($\Lambda$)
are compared for different parameters $\lambda^2 \eta$ at $t=\tau_s$.

\begin{thebibliography}{99}
\bibitem{Lloyd} S. Lloyd, Science {\bf 261}, 1589~(1993).
\bibitem{Bennett} C.H. Bennett, Phys. Today {\bf 48}, 24~(1995).
\bibitem{Jozsa94} R. Jozsa and B. Schumacher, J. Mod. Opt. {\bf 41},
2343~(1994).
\bibitem{Schumacher95} B. Schumacher, Phys. Rev. A {\bf 51}, 2738~(1995).
\bibitem{Shor95} P.W. Shor, Phys. Rev. A {\bf 52},2493~(1995).
\bibitem{Steane} A. Steane, Phys. Rev. Lett. {\bf 77}, 793~(1996).
\bibitem{Laflamme} R. Laflamme, C. Miquel, J.P. Paz, and W.H. Zurek, Phys.
Rev.Lett. {\bf 77}, 198~(1996).
\bibitem{Nielsen} B. Schumacher and M.A. Nielsen, Phys. Rev. A {\bf 54},
2629~(1996).
\bibitem{Jozsa95} R. Jozsa, J. Mod. Opt. {\bf 41}, 2315~(1995).
\bibitem{Wiesner} C.H. Bennett and S.J. Wiesner, Phys. Rev. Lett. {\bf 68},
3121~(1992).
\bibitem{Brassard} C.H. Bennett, G. Brassard, C. Cr{\'e}peau, R. Jozsa, A.
Peres, and W.K. Wootters, Phys. Rev. Lett. {\bf 69}. 2881~(1992).
\bibitem{Shor94} P. Shor, Proc. 35th Annual Symp. on the Foundations of
Computer Science (IEEE Press, Los Alamitos, 1994), p.124.
\bibitem{DiVincenzo} D.P. DiVincenzo.Science {\bf 270},255~(1995).
\bibitem{Barenco} A. Barenco, D. Deutsch, A. Ekert, and R. Jozsa, Phys. Rev.
Lett. {\bf 74}, 4083~(1995)
\bibitem{Loss} D. Loss and D.P. DiVincenzo, Phys. Rev. A {\bf 57},
120 (1998).
\bibitem{Schumacher96} B. Schumacher, Phys. Rev. A {\bf 54},2614~(1996).
\bibitem{Reichl} L.E. Reichl, A Modern Course in Statistical Physics
(University of Texas Press, Austin, 1980).
\bibitem{Tokuyama} M. Tokuyama and H. Mori, Prog. Theor. Phys. {\bf 55},
411~(1975). 
\bibitem{Hashitsume} N. Hashitsume, F. Shibata, and M. Shingu, J. Stat. Phys.
{\bf 17}, 155~(1977).
\bibitem{Saeki82} M. Saeki, Prog. Theor. Phys. {\bf 67}, 1313~(1982).
\bibitem{Saeki86} M. Saeki, J. Phys. Soc. Jpn. {\bf 55}, 1846~(1986).
\bibitem{Ahn94}D. Ahn, Phys. Rev. B {\bf 50}, 8310~(1994).
\bibitem{Ahn95}D. Ahn, Phys. Rev. B {\bf 51}, 2159~(1995).
\bibitem{Ahn97}D. Ahn, Prog. Quantum Electron. {\bf 21}, 249~(1997).
\bibitem{Ahn98}D. Ahn, IEEE J. Quantum Electron. {\bf 34}, 344~(1998).
\bibitem{Nakajiama} S. Nakajiama, Prog. Theor. Phys. {\bf 20}, 948~(1958).
\bibitem{Zwanzig}R. Zwanzig. J. Chem. Phys. {\bf 33}, 1338~(1960).
\bibitem{Hubb}N. W. Ashcroft and N. D. Mermin, Solid State Physics
(Saunders, Philadelphia, 1967), Chap. 32.
\bibitem{Poyatos} J.F. Poyatos, J.I. Cirac, and P. Zoller, Phys. Rev. Lett.
{\bf 78}, 390~(1997).
\bibitem{Palma} G. M. Palma, K. Suominen, and A. K. Ekert, {\it preprint}, (1999).
\end{thebibliography}
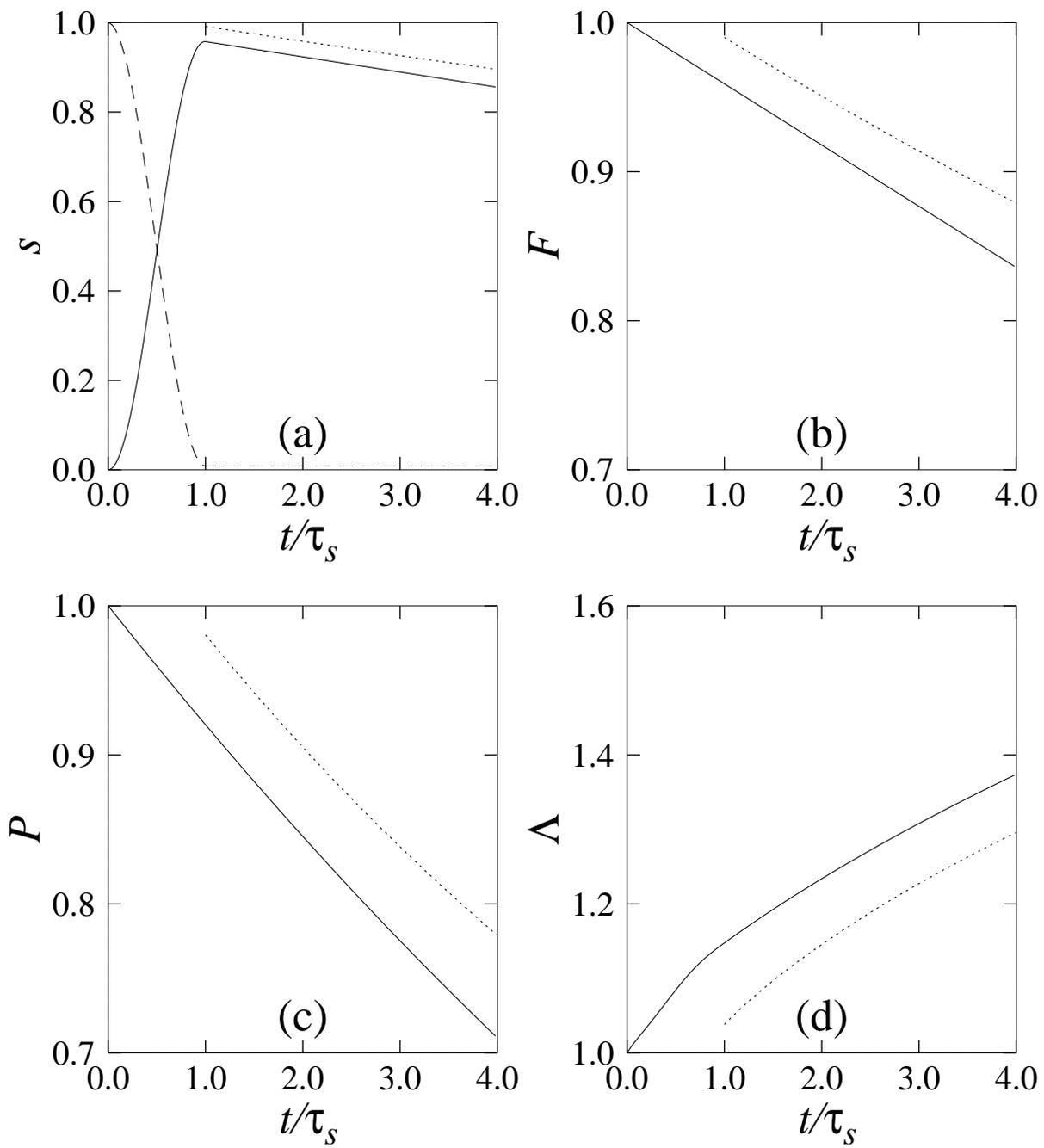
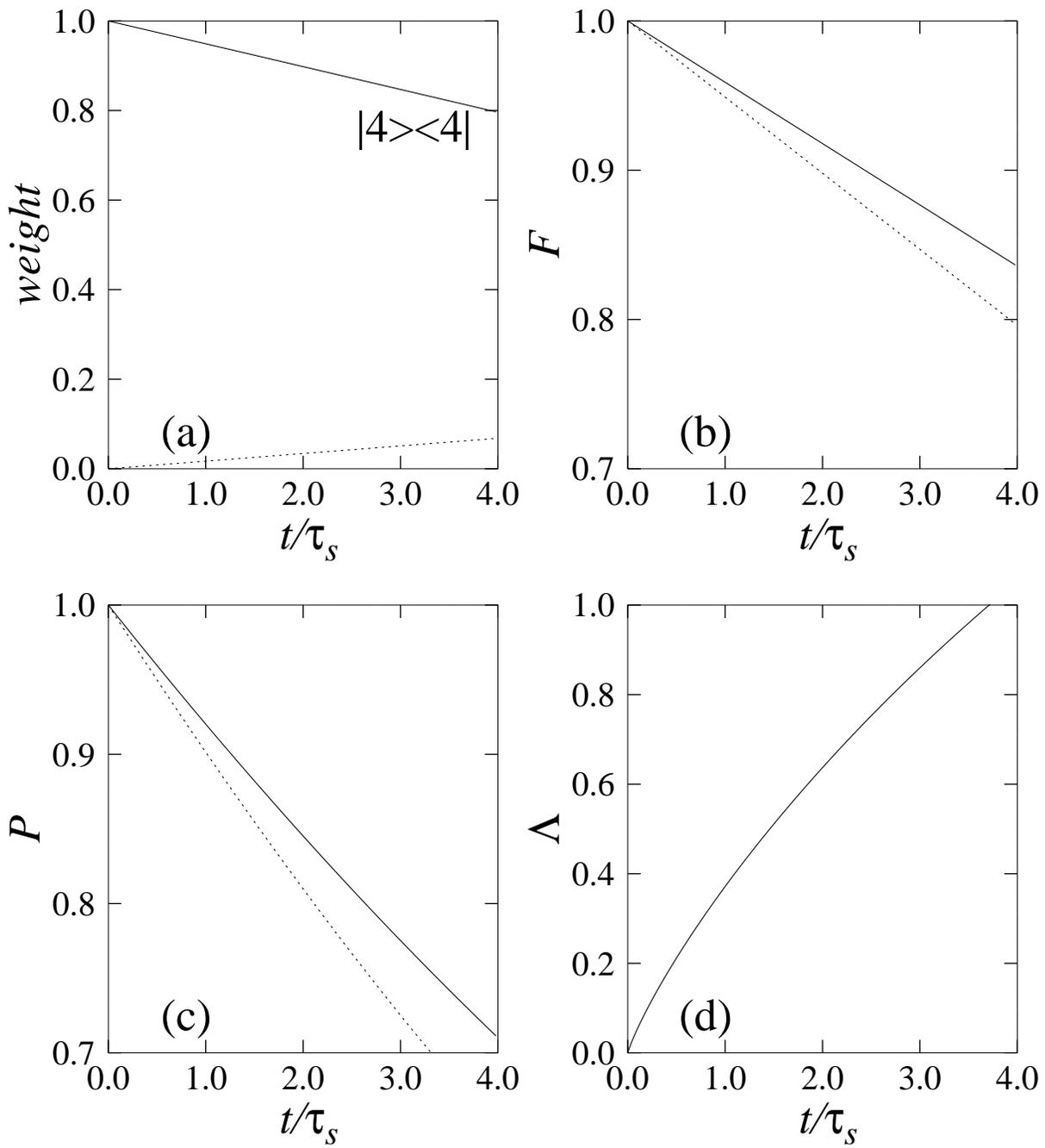
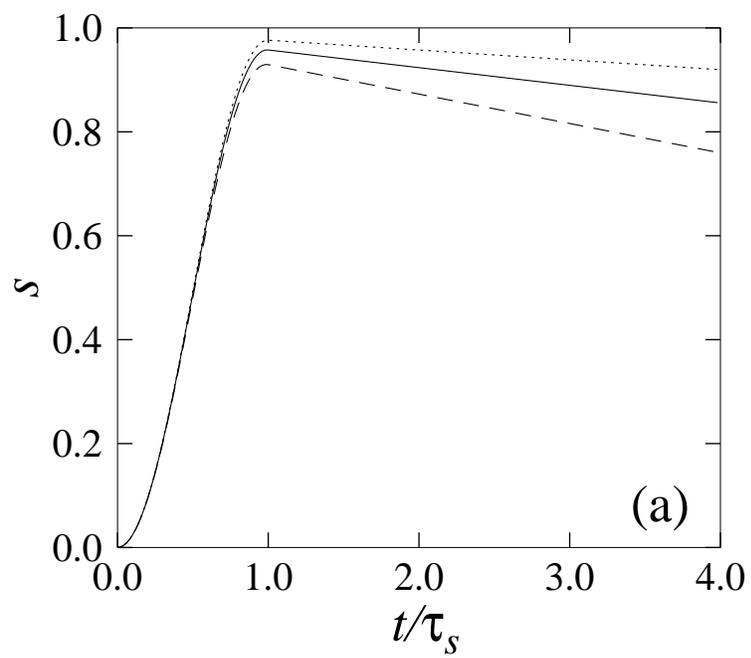
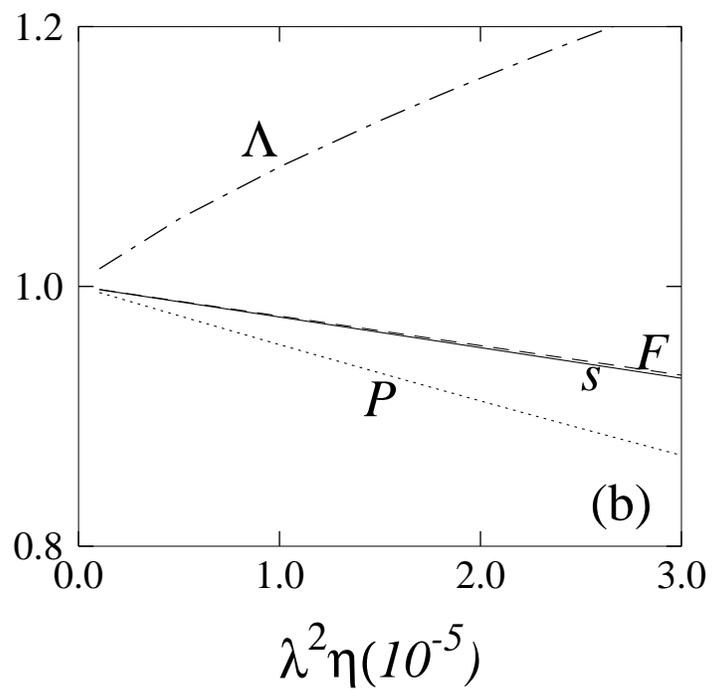
\end{document}